\documentclass[aps,prl,twocolumn,floatfix,amsmath,showpacs]{revtex4-1}
\usepackage{natbib}
\usepackage{color}
\usepackage{graphicx}
\usepackage{amsfonts}
\DeclareGraphicsExtensions{.pdf,.png,.jpg}

\begin{document}

\newcommand{\wq}[1]{\textcolor{blue}{#1}}

\title{Intermediate Range Structure in Ion-Conducting Tellurite Glasses
}
\author{M.~A.~Frechero}
 \affiliation{Secci\'on Fisicoqu\'{\i}mica, INQUISUR-UNS-CONICET and 
Departamento de Qu\'{\i}mica, Universidad Nacional del Sur, 
Avenida Alem 1253, 8000 Bah\'{\i}a Blanca, Argentina}
\author{L.~Padilla}
%\email{lorenapadilla.r@gmail.com}
\author{H.~O.~M\'artin}
\author{J.~L.~Iguain}
\email{iguain@mdp.edu.ar}
\affiliation{Instituto de Investigaciones F\'{\i}sicas de Mar del Plata (IFIMAR) and
Departamento de F\'{\i}sica FCEyN,\\
Universidad Nacional de Mar del Plata, De\'an Funes 3350, 7600 Mar del
Plata, Argentina}

\begin{abstract}
We present ac conductivity spectra of tellurite  
glasses at several temperatures. For the first time, we report  oscillatory modulations
at frequencies around MHz. This effect is more pronounced the lower the temperature, and 
washes out when approaching the glass transition temperature $T_g$. 
We show, by using a minimal model, how this modulation may be attributed to the fractal structure of
the glass at intermediate mesoscopic length scales.

\end{abstract}
\maketitle

Over the last 30 years, the study of the atomic structure of inorganic 
glasses has revealed a very rich order both from a local point of view 
(individual polyhedron coordination), as at intermediate range 
(the way polyhedra are connected among them). Of course, 
it keeps
long range disorder, as all of us interpret glassy materials.
Many experiments have exposed a quasi-periodicity in the network, 
including a kind of channels of network modifiers, in
a direct relationship with ionic transport mechanisms.
However, the comprehension of the relaxation laws in these 
materials is still incomplete, and many  
questions remain unanswered~\cite{Salmon2002, *Greaves2007,*Sanda2007}. 

Conductivity spectroscopy is a widely employed technique to investigate
ion dynamics at different time and length scales~\cite{Funke1997}, 
and considerable effort has been dedicated to
obtain an universal behavior bringing the spectra to collapse on a single
curve.
A scaling law without arbitrary parameters
was first proposed in 
Ref.~\cite{Roling1997} for single ion conducting glasses, and
then slightly modified to account for glasses with
a wide variation of alkali content~\cite{Sidebottom1999}.
However, that the shape of the spectra is not universal but depends on glass 
composition was later shown in Ref.~\cite{Roling2000}. 

For one given compound, a simple scaling law can be obtained on the basis of
the time-temperature superposition principle (TTSP), stating that,
at a temperature $T$, only one characteristic time scale exists:
the crossover time between sub-diffusive and normal diffusive ion behavior. 
The form
\begin{equation}
{\sigma(\nu)}/{\sigma_{dc}}=F\left({\nu}/{\nu_0}\right),
\label{ttsp}
\end{equation}
for the conductivity spectra $\sigma(\nu)$, expresses the TTSP in frequency space,
with $\sigma_{dc}$ the
dc conductivity, $\nu_0$ the crossover frequency (both temperature dependent),
and $F$ the scaling function (which depends on the compound).

In 1985, Summerfield proposed that $\nu_0=\sigma_{dc}T$ for
 amorphous semiconductors~\cite{Summerfield1985, *Balkan1985}. 
This assumption has also been shown to be valid on some 
single ion conducting glasses~\cite{Roling1997,Sidebottom1999,Roling2000,Zielniok2008},
but cannot be applied generally. The conductivity spectra fail to collapse as 
suggested by Summerfield not only for various mixed alkali 
glasses~\cite{Cramer2002,*Imre2002} but also for single ion tellurite 
glasses~\cite{Murugavel2002}. In the latter case the 
universal form of the spectra is obtained by adopting $\nu_0=\sigma_{dc}T/T^\alpha$,
at the cost of introducing a new parameter $\alpha$. More recently, this kind
of universality has been demonstrated for the conductivity spectra of polyelectrolyte
complexes~\cite{Imre2009}.

In this Letter, we report an oscillatory behavior observed
in electrical conductivity measurements on tellurite glasses 
at frequencies around $1~$MHz. These glasses have a mainly ionic 
conductivity behavior, with the 
alkaline cations as charge carriers.
 Generally, 
at frequencies above $100$ Hz and  temperatures bellow the 
glass transition temperature $T_g$, there is little dispersion 
in the electrical conductivity and the response of the system 
is related to the immediate vicinity of the alkali cation.
The presence of an oscillating modulation is a clear indication of
the existence of several relevant length scale in the 
ion diffusion problem, and that simple scaling forms 
as Eq.~(\ref{ttsp}) cannot be applied. 
We propose an interpretation of this phenomenon based on the
structure of the glass at intermediate mesoscopic length scales.

{\bf Experimental- }
The samples were prepared by a standard melt quenching technique from initial 
mixtures of proper quantities of components ($99.99\%$ pure): TeO$_2$, 
V$_2$O$_5$, MoO$_3$, and Na$_2$CO$_3$ or Li$_2$CO$_3$. 
The amorphous character of each resulting
 solid, \mbox{$0.6$Na$_2$O-$0.4$[0.5V$_2$O$_5$0.5MoO$_3$]-2TeO$_2$}, 
with \mbox{$T_g=535.7K$}, and
\mbox{$0.6$Li$_2$O-$0.4$[0.5V$_2$O$_5$0.5MoO$_3$]-2TeO$_2$} with $T_g=539.5K$, 
was tested by 
X-ray diffraction analysis and confirmed by the 
Differential Scanning Calorimetry (DSC). Glass disks of thickness ranging 
between $0.5-1.0$ mm, were cut from the obtained cylinder and polished with 
very fine quality lapping papers. The electrodes for electrical measurements 
were made using silver conducting paint to which metallic leads were attached. 
 The conductivity of each sample $\sigma$ was determined, as a function of the frequency $\nu$,  
by standard a.~c.~impedance 
methods using a Novocontrol BDS-80 system. We have explored the electrical
response at three distinct values of the temperature $T$: $223$K, $353$K and 
$473$K.
The applied voltages $V(t)=V_{a}\cos(2\pi\nu t)$ with a small 
amplitude $V_{a} = 100$ mV ensured a linear response throughout the frequency 
 range. 

\begin{figure}
\includegraphics[width=.85\linewidth]{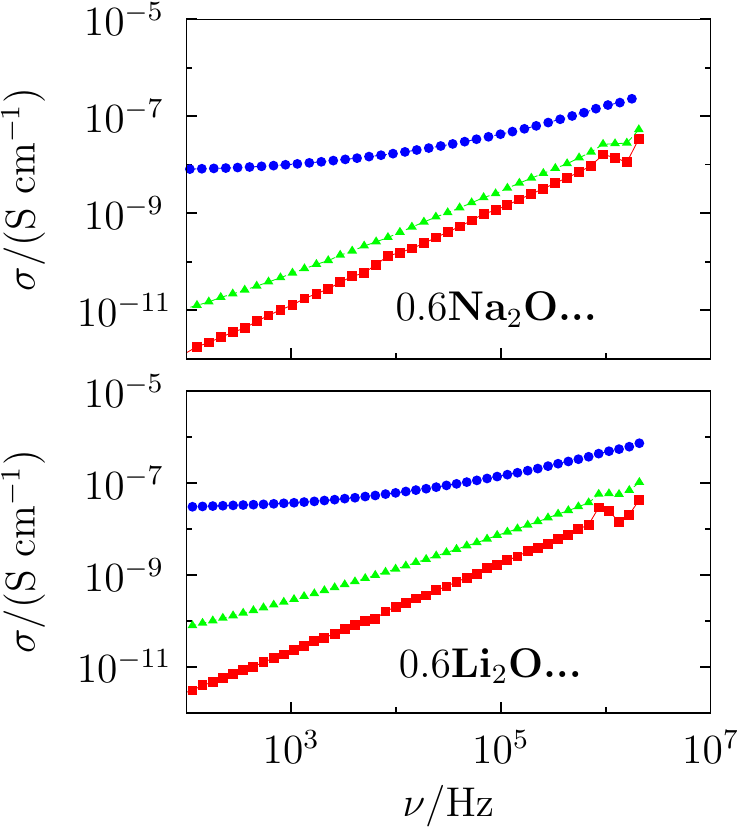}
\caption{
(Color online) Experimental a.~c.~electrical conductivities as a function of the frequency
at three different temperatures: 223K (red squares), 353K (green triangles), 473K 
(blue circles). Top panel: \mbox{0.6Na$_2$O-0.4[0.5V$_2$O$_5$0.5MoO$_3$]-2TeO$_2$} compound. 
Lower panel:
\mbox{0.6Li$_2$O-0.4[0.5V$_2$O$_5$0.5MoO$_3$]-2TeO$_2$} compound.
Errors are smaller than symbol sizes}
\label{experimental}
\end{figure}
 
The experimental conductivity spectra of the obtained compounds
are plotted in Fig.~\ref{experimental}.  
A simple inspection of this figure reveals that, as the frequency increases, 
$\sigma(\nu)$ shows not only the usually expected crossover from a nearly 
constant conductivity (out of the plot scale for the lower temperatures), 
to a  power-law behavior, but also that an oscillatory modulation appears at 
larger frequencies ($\nu\sim 1$MHz). Note also that the amplitude of the
oscillations decreases with  the increasing of temperature, and
becomes negligible for high enough $T$.

{\bf Interpretation- }
Ion conducting glasses can be considered as weak electrolytes where a density 
$\rho$ of ions, which depends on temperature, becomes mobile
and explore the sample in a random walk (RW)~\cite{Souquet2010}.
Two basic hopping mechanisms have been proposed~\cite{Greaves1995}. 
In the first, the so called {\it network hopping}, the diffusing ion moves 
by jumps in a random network 
 consisting only on bridging oxygens (BO). These hoppings  involve little 
conformational change in the network, and have a relatively small activation 
energy $\Delta E_{\tt net}$ ($\approx 0$).
In the second mechanism, referred to as {\it intrachannel hopping}, the alkalies 
are mainly coordinated to non-bridging oxygens (NBO), and mobile ions diffuse 
in a network of 
conducting channels, which results of connecting NBO's together. 
These jumps do entail important modifications in the local NBO and BO 
configurations, and a comparatively high activation energy $\Delta E_{\tt ic}$ 
($\gg \Delta E_{\tt net}$) results.
Though there exist materials where the ionic transport can be represented 
with one of these two ideal networks (compensated aluminosilicates, the first; 
silicates, the second), in general, both mechanisms are present.   
\begin{figure}
\includegraphics[width=.8\linewidth]{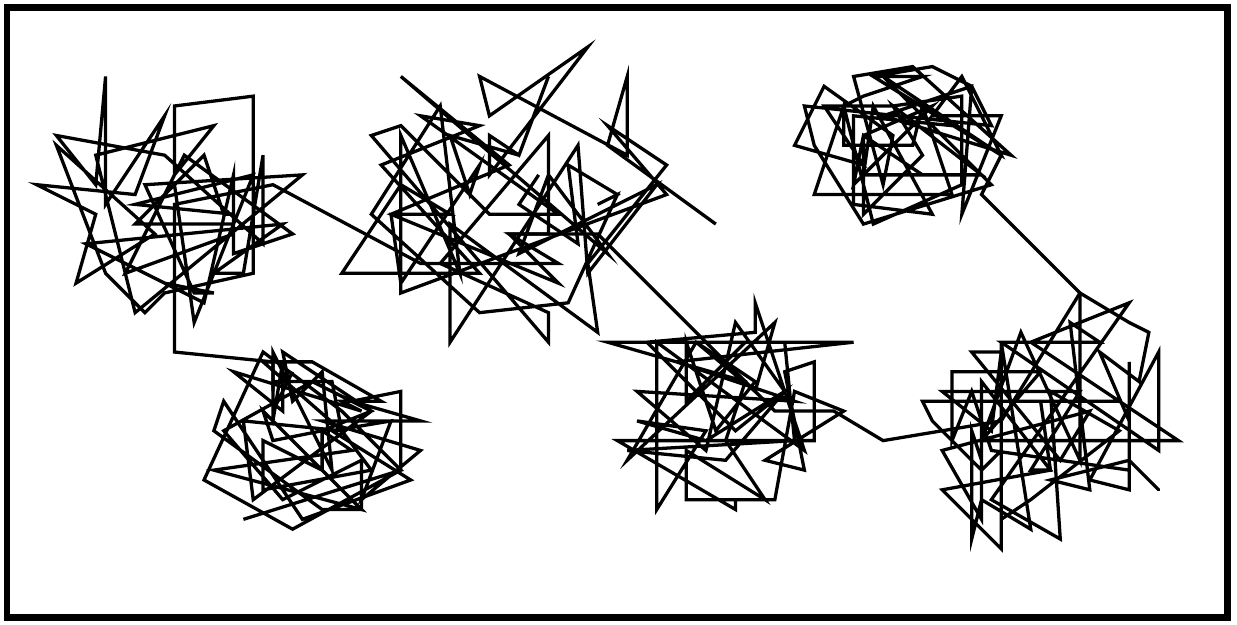}
\caption{
Sketch of single ion diffusion in an oxide glass. At short distances, 
the behavior is dominated by hoppings in a network of BO's only; with a very small activation 
energy $\Delta E_{\tt net}$. When this network does not percolate the sample, 
jumps among NBO's, with a high activation energy $\Delta E_{\tt ic}$
($\gg \Delta E_{\tt net}$), play a role.    
}
\label{esque-difu}
\end{figure}
 
In the first stages of the diffusive motion, ion
migration occurs mainly by network hopping due to energetic reasons. However, if the random network of BO's 
does not percolate the sample but consists on disconnected regions, after a 
typical time $t_1$, the ion will reach the border of one of these regions, 
the diffusion will slow down, and the particle will behave as if trapped 
in a cage and with an escape time $\tau$. In addition, if the typical linear size 
of a cage is not so large, $\tau\sim \exp(E_{\tt ic}/k_BT)$ ($k_B$ is the Boltzmann constant), 
and we expect the behavior of mobile ions sketched in Fig.~\ref{esque-difu}.
Thus, at intermediate times
$t$ ($t_1<t<\tau$), the ion mean-square displacement behaves as 
$\Delta^2r(t) \sim t^{\eta}$, with a very small RW exponent ($\eta \simeq 0$),
because of the effect of the cages.
For times longer than $\tau$, the ions become 
able to explore the whole sample, and normal diffusion results ($\eta\simeq 1$)
if large-scale homogeneity is assumed.

According to the linear response theory, the conductivity spectrum
 is related to the ion mean-square displacement by

\begin{equation}
\sigma(\nu)=-2\pi^2\nu^2\frac{\beta\rho e^2}{dH_R}\lim_{\epsilon\rightarrow 0^+}
               \int_0^\infty\!\!\!\! dt\, \Delta^2 r(t) \cos(2\pi\nu t) e^{-\epsilon t},
\label{tf}
\end{equation}
where $\beta=1/k_BT$, $d$ is the substrate 
dimension,  $e$ is the ion charge, and 
$H_R$ is the Haven ratio, which expresses the multi-particle 
correlations~\cite{Havlin,*Heuer2002}. A simple dimensional analysis of 
Eq.~(\ref{tf}) leads to  

\begin{equation}
 \Delta^2r(t) \sim t^{\eta} \Leftrightarrow \sigma(\nu)\sim 
\nu^{1-\eta},
\label{expos}
\end{equation}
and then, the above mentioned change in the kinetics 
of diffusion is reflected in 
the behavior of the function $\sigma(\nu)\sim \nu^\zeta$, 
which, as the frequency increases, 
undergoes the crossover  $\zeta\simeq 0\longrightarrow \zeta\simeq 1$. 

Long-time or short-frequency ion behavior  
can be captured by modeling the substrate as a set of cells of convenient size, 
separated by energetic barriers of height $\Delta E_{\tt ic}$. 
However, to understand the mechanisms responsible for the oscillations 
appearing at higher frequencies, we have to watch inside these cells, 
at distances larger than atomic radius but smaller than cell typical size.  
In what follows, we show that the experimental results in 
Fig.~\ref{experimental} can be qualitatively reproduced if
we assume that the BO network has a fractal structure,
similar to the well-known Vicsek model~\cite{vic}.

For a single particle diffusing on a self-similar substrate, 
a hierarchical set of length-dependent diffusion coefficients $\{D^{(n)}, n\in \mathbb{N}\}$ exists, 
with $\{D^{(n)}/D^{(n+1)}=1+\lambda, n \in \mathbb{N}\}$, and
where $\lambda>0$ is a constant; meaning that
the RW mean-square displacement behaves as 
$\Delta^2r(t)\sim D^{(n)}t$, for $\sqrt{\Delta^2r}$ in the range $(L^n, L^{n+1})$, 
where $L$ is the basic length of the fractal. 
The decreasing of the diffusion coefficient with the increasing of the
length scale produces an oscillating modulation in $\Delta^2r(t)$~\cite{Padilla2010}, and, according to  
Eq.~(\ref{tf}), also in $\sigma(\nu)$. Note that oscillations 
in the mean-square displacement at short times imply oscillations in the conductivity at high frequencies.

\begin{figure}
\includegraphics[width=.65\linewidth]{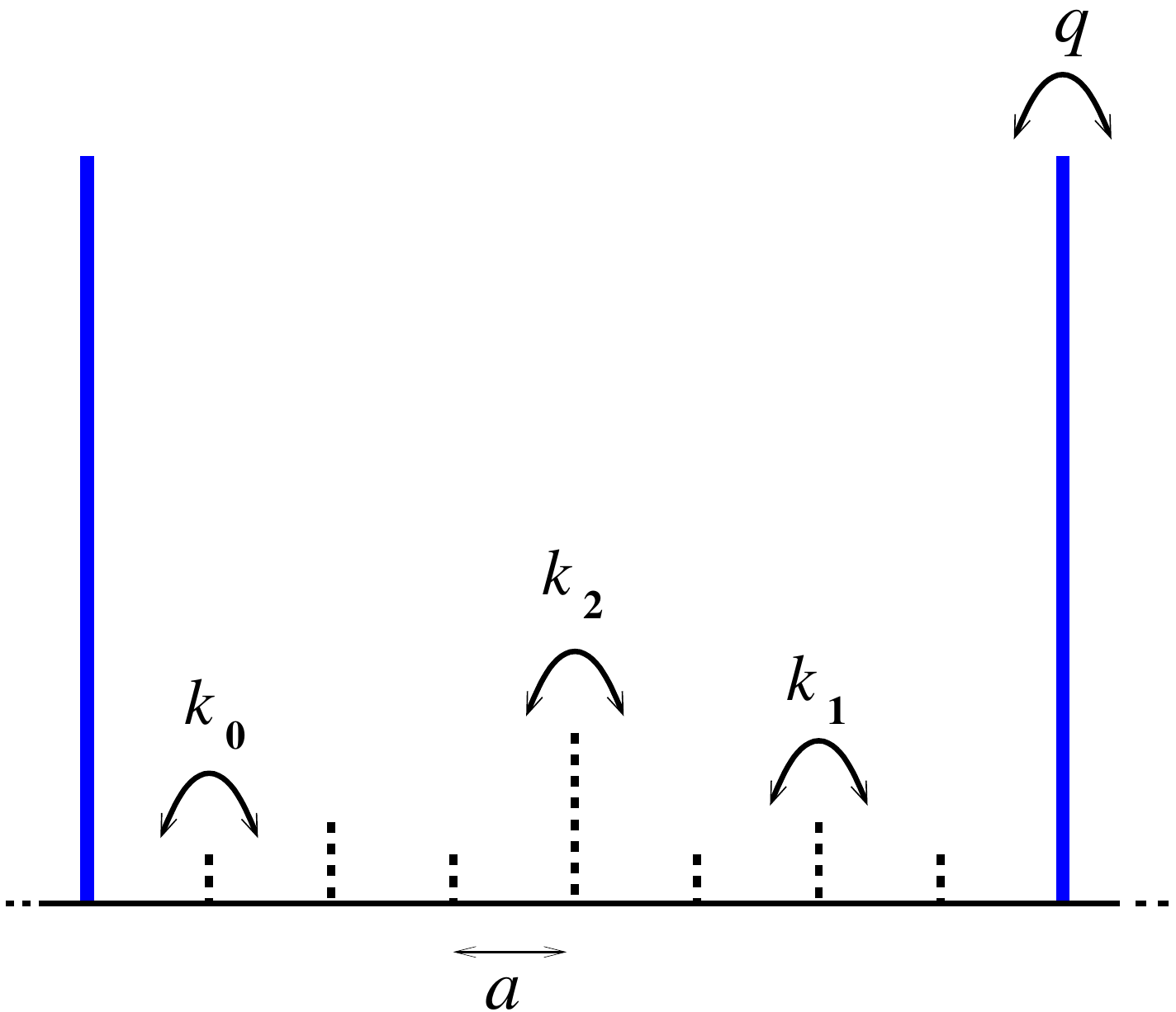}
\caption{A one-dimensional unit cell. The topological hopping rates $k_i$ ($i=0,1,2$)
represent network hoppings. They are built in a self-similar manner and do not depend
on temperature. The energetic hopping rate $q=k_0\exp(-E_{\tt ic}/k_BT)$ 
accounts for intrachannel jumps.
}
\label{model}
\end{figure}

The most relevant aspects of RW in a self-similar structure
can be effectively captured by a one-dimensional model, in which
$L$ and $\lambda$ are introduced as parameters~\cite{Padilla2009}.
Thus, to keep our approach simple we consider a particle moving on a periodic 
lattice in one dimension, with the unit cell as sketched in Fig.~\ref{model}. 
At every time step the particle can hop to a nearest
 neighbor lattice site. The hopping rates depend on the initial and 
final sites only and are represented in the drawing by vertical segments
or barriers. Two kinds of barriers exist. 
The \emph{topological} barriers (dashed segments)
correspond to the network hopping processes. Their rates, which do not 
depend on temperature and decrease with the length scale, 
were obtained from the minimal self-similar model in Ref.~\cite{Padilla2009}. For concreteness, 
we have chosen $L=2$ and $1+\lambda=20$,
leading to $k_1/k_0=2.56\times 10^{-2}$ and $k_2/k_0=6.41\times 10^{-4}$.
The \emph{energetic} barriers 
(full segments) correspond to the 
intrachannel hopping processes occurring at a rate 
$q=k_0\exp (-E_{ic}/k_BT)$. 

\begin{figure}
\includegraphics[width=.75\linewidth]{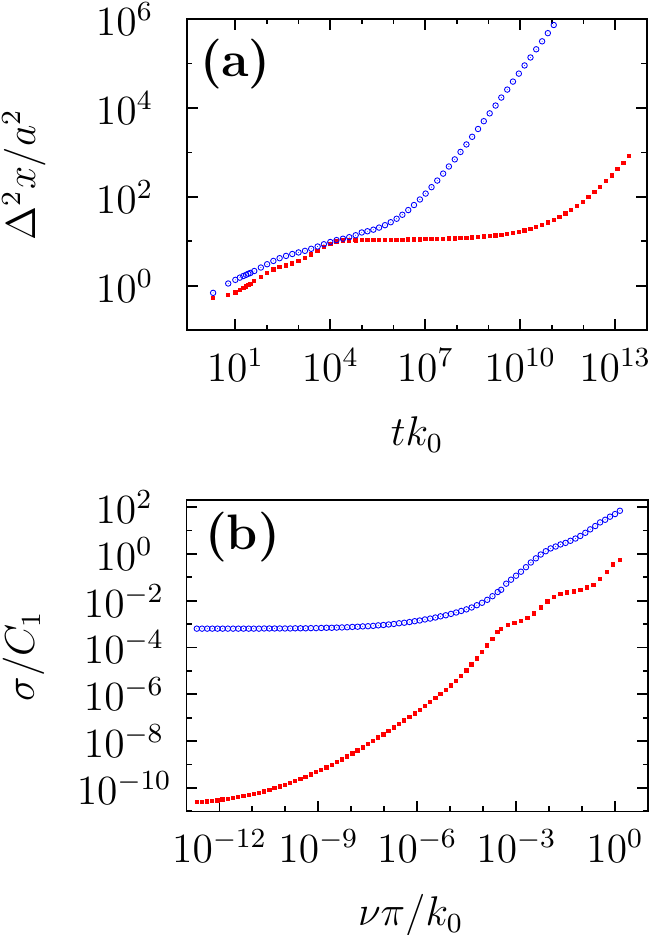}
\caption{(Color online) (a)~Mean-square displacement as function of time 
for the one-dimensional minimal model at two temperatures $T_1$ (red-lower symbols)
and $T_2=2T_1$ (blue-upper symbols). 
(b)~Conductivity spectra calculated with the data in (a).
}
\label{simu_L2}
\end{figure}

Let us assume that at a low enough temperature $T_1$  the structure
where mobile ions diffuse can be modeled by a periodic array of the unit
cells in Fig.~\ref{model}. We fix $T_1$  by asking that the corresponding 
intrachannel transition rate is $q(T_1)=10^{-12}k_0$.
For this substrate, the RW mean-square displacement as a function of time, obtained
by Monte Carlo simulations, is shown
in Fig.~\ref{simu_L2}-a (lower data points). The results were averaged over
$10^4$ independent runs with random initial positions.
Three different regimes can
be clearly observed.
At short times, $\Delta^2r(t)$ exhibits a sub-diffusive power-law behavior
modulated by log-periodic oscillations due to the self-similar structure 
of the unit cell.
At intermediate times, it shows a plateau, caused by 
the constraint in diffusion imposed  by the 
energetic barriers.
Finally, at long enough times ($\gg 10^{12}/k_0$), $\Delta^2r(t)\sim t$.
As explained in Ref.~\cite{Padilla2009}, the crossovers among the length scales 
$\ell_n=a 2^n$, for $n=0, 1,$ and $2$, cause a sub-diffusive behavior modulated by oscillations.
At distances larger than 8$a$, the substrate can be considered as periodic, and the diffusion 
becomes normal.

At a higher temperature, we consider the possibility of some level of 
randomness in the BO network, which we introduce in the model by  
shuffling 
the locations of the topological barriers. 
Although the so-obtained substrate will depend on the probability distribution 
function involved in the shuffle procedure, it has been shown that, 
in the limit of full randomization,
the oscillating modulation is washed out but the RW exponent remains as in
the self-similar substrate~\cite{Padilla2011}. 
By using the MC protocol described above, we have studied the RW of a 
single particle at a temperature $T_2=2T_1$, giving an
 intranetwork hopping rate ($k_0$ units) 
\mbox{$q(T_2)/k_0=\left(q(T_1)/k_0\right)^{1/2}= 10^{-6}$}. We have
used a substrate in which the topological barriers 
are randomized in the $30\%$ of the 
cells (the other $70\%$ remain ordered as in Fig.~\ref{model}). 
These values are arbitrary but useful to appreciate
the trend of the effects of randomness.
The corresponding numerical RW mean-square displacement is
 plotted as a function of time in Fig.~\ref{simu_L2}-a (upper data points). 
As compared with the behavior at $T_1$, the amplitude of the oscillations 
decreases because of randomness, and the normal diffusion regime is reached 
at a shorter time due to the increasing of the rate $q$.
 
Using the data in Fig.~\ref{simu_L2}-a, we have calculated the corresponding electrical
conductivities, by numerical integration of Eq.~(\ref{tf}),
which can be rewritten as

\begin{equation}
\sigma(\nu;\beta)=-C_1 \left({\beta\rho}/{\beta_1\rho_1}\right)
I(\nu;\beta)\;.
\label{tf2}
\end{equation}
Here $\rho_1$ is the carrier density at the inverse temperature 
$\beta_1=1/k_BT_1$, $C_1$ is a reference conductivity 
\mbox{$C_1=2e^2k_0a^2\beta_1\rho_1 /H_R$}, and 
$I(\nu;\beta)$ is the non-dimensional function
\begin{equation}
I(\nu;\beta)=\lim_{\epsilon\rightarrow 0^+}
 \int_0^\infty \!\!\!\!\! dt\, k_0\,\left(\frac{\nu\pi}{k_0}\right)^2 
\frac{\Delta^2 r(t;\beta)}{a^2} \cos(2\pi\nu t) 
e^{-\epsilon t}.
\label{inte}
\end{equation}

The density of mobile ions depends strongly on temperature. 
It has
been reported that $\rho$ increases in a factor
of around $10^2$ when $T$ is doubled from $200 K$ to $400 K$ 
\cite{Souquet2010}. As we want to relate $T_1$ with the lowest temperature
of our measurements (see Fig.~\ref{experimental}),
we use a value of $50$ as a good estimate of the factor in parentheses in Eq.~(\ref{tf2}), 
when $T_1\simeq 200 K$, and $\beta=\beta_2=\beta_1/2$. We neglect the possible variation of $H_R$
with the temperature.
The numerical conductivity spectra are plotted in Fig.~\ref{simu_L2}-b. 
Let us remark the good qualitative agreement between model and experiment.

In summary, we report for the first time the 
experimental ac conductivity spectra of tellurite glasses showing oscillatory 
modulations at frequencies around MHz, which reflect a non-trivial structure
of the glass at mesoscopic distances. In order to account for these 
oscillations, we introduce a minimal model describing ion diffusion in a 
one-dimensional substrate with a short-length fractal  structure;
self-similar at low enough $T$, and fully disordered for $T\lesssim T_g$.
We find a very good qualitative agreement between theoretical and experimental
results, which supports our hypothesis that a fractal structure exists
in oxide glasses at intermediate length scales.

{\bf Acknowledgments-} We are grateful to the Grupo de F\'{\i}sica de Materiales Complejos - UCM
for kindly sharing their Novocontrol BDS-80 system. This work was supported by 
UNMdP and CONICET (PIPs 041 and 431).

\end{document}